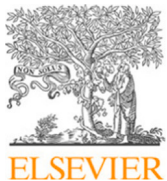
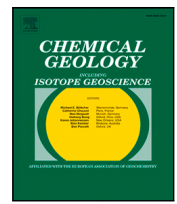

Research paper

# Water solubility in silicate melts: The effects of melt composition under reducing conditions and implications for nebular ingassing on rocky planets

Maggie A. Thompson [a,b,*,1], Paolo A. Sossi [a], Dan J. Bower [a], Anat Shahar [b], Christian Liebske [a], Julien Allaz [a]

[a] *Institute for Geochemistry and Petrology, Department of Earth Sciences, ETH Zürich, Clausiusstrasse 25, Zürich, 8092, Switzerland*
[b] *Earth and Planets Laboratory, Carnegie Institution for Science, Washington, DC, 20015, USA*

## ARTICLE INFO

## ABSTRACT

Rocky planet atmospheres form and evolve through interactions between the planet's surface and interior. If a growing rocky planet acquires enough mass prior to the dissipation of the nebular gas disk, it can gravitationally capture a 'primary' atmosphere dominated by $H_2$. At the same time, these young, rocky bodies are likely to have partial or global magma oceans as a result of the heat from accretion, core formation and radioactive decay of short-lived major element isotopes. During this magma ocean stage, the dissolution of volatile, life-essential elements, such as hydrogen in the form of water or $H_2$, into the magma is critical in determining the extent to which a rocky planet can maintain these elements over time. However, our ability to quantify the amount of hydrogen dissolved in the magma oceans of rocky planets is limited by the lack of experimental constraints on H-bearing species' solubilities at relevant pressure and temperature conditions, including those expected for the early Earth. Here we experimentally determine the solubility of water in silicate melts of various compositions in the Ca–Mg–Al–Si–Fe–O system at a total pressure of 1 bar and temperatures from 1673–1823 K, synthesized in a $H_2$–$CO_2$ gas-mixing furnace. We explored a range of $H_2$ and $H_2O$ fugacities, with $f(H_2)$ ranging from 0.68 to 0.98 bar and $f(H_2O)$ from 0.02 to 0.16 bar. The amounts of dissolved water were quantified by transmission Fourier-Transform infrared spectroscopy (FTIR) analysis, using the intensity of the 3550 cm$^{-1}$ absorption feature corresponding to the OH stretching band. We detect water dissolution as $OH^-$ and find that the amount of dissolved $OH^-$ increases with $f(H_2O)^{0.5}$. We use Bayesian parameter estimation to derive a robust water solubility law that includes the effects of melt composition and temperature. Using this solubility law, we estimate that ∼100 ppm of hydrogen can dissolve into a 1 $M_{Earth}$ planet with a surface pressure of ∼300 bars set by accretion of solar-like nebular gas, matching current estimates of the hydrogen content of the bulk silicate Earth. However, multiple lines of evidence, including Earth's core formation age post-dating dissipation of the solar nebula, indicate that nebular ingassing did not contribute significantly to Earth's present-day hydrogen budget. Nevertheless, for rocky planets in general, ingassing of a primary atmosphere may be an important source and initial storage mechanism for hydrogen-bearing species in a planet's interior, provided it grew to a sufficient mass within the lifetime of the solar nebula.

## 1. Introduction

During planet formation, the heat from accretion and impacts likely generates global magma oceans on planetary embryos and young planets (Chao et al., 2021). Chemical equilibrium between magma oceans and atmospheres will occur if the transport of material between the interior and atmosphere is efficient. For both an Earth-sized molten mantle, and on planetary embryos (∼0.05–0.2 $M_\oplus$), mixing is fast relative to the atmospheric cooling, and therefore equilibrium is readily achieved (Sossi et al., 2023; Young et al., 2023). The atmosphere that forms during the magma ocean stage is important for setting a planet's mantle volatile inventory, atmospheric stability, and habitability potential (Catling and Kasting, 2017). Owing to chemical equilibrium between the magma ocean and atmosphere, the atmospheric composition depends primarily on two factors: (1) the planet's bulk inventory of atmosphere-forming elements (i.e., H, C, N, O, S) and (2) the solubilities of these gases in the magma ocean. The solubilities of gas species

---

depend on several factors including the surface pressure, temperature and the melt composition.

Solubilities are largely determined experimentally and much of the existing work has focused on those of the major volatile species (e.g., $H_2$, $H_2O$, $CO_2$, $CH_4$, $SO_2$) in silicate melts under moderate to high-pressure conditions (hundreds of bars to several GPa) relevant for Earth's crust and mantle (e.g., Dixon et al., 1995; Hirschmann et al., 2012; Boulliung and Wood, 2022). Because, all else being equal, the mole fraction of a dissolved volatile species is proportional to the partial pressure (or, more precisely, the fugacity) of the conjugate vapor species in the gaseous phase or its fugacity raised to some power (e.g., 0.5 for $H_2O$ Hamilton et al., 1964), solubility laws determined by fits to experiments conducted at low total pressures (~1-200 bar) are sparse, despite their relevance for terrestrial planets. Indeed, the atmospheric pressure above Earth's magma ocean is expected to have been below several hundred bars, assuming the bulk volatile content during Earth's magma ocean stage is similar to that observed today (Sossi et al., 2020; Gaillard et al., 2022).

Existing studies on hydrogen-bearing species' solubilities have primarily focused on Earth-like silicate melts (e.g., basalt, andesite, albite, rhyolite). Early works in the 1960-70s investigated hydrogen solubility and diffusivity in fused silica under a range of temperatures (~500-1100 °C) and pressures (~1 bar to several kbar) (e.g., Doremus, 1966; Bell et al., 1962). These earlier studies note that hydrogen can dissolve both "physically" (i.e., as molecular $H_2$ and $H_2O$, without chemical reaction with the glass, residing in interstitial spaces) and "chemically" (i.e., as $OH^-$ groups, via chemical reaction with components of the melt or glass). The speciation of dissolved hydrogen in melts under various conditions and its change upon quenching into a glass remain active areas of study (Zhang and Ni, 2010).

In general, low-pressure (< kbar) studies find that water dissolves as $OH^-$ while higher-pressure studies also detect molecular $H_2O$ in the glass (we note, however, that the species in the corresponding liquid is always $OH^-$, with $H_2O$ forming upon quenching Dingwell and Webb, 1990). At high $pH_2$, molecular $H_2$ in glasses is detected (Bell et al., 1962; Luth et al., 1987; Faile and Roy, 1971; Persikov et al., 1990; Gaillard et al., 2003; Hirschmann et al., 2012). For example, Gaillard et al. (2003) determined $H_2$ solubility in rhyolitic obsidian melt from 0.02-70 bar, showing that molecular $H_2$ was detected in the glass only at the highest pressures (Gaillard et al., 2003). Persikov et al. (1990) and Luth et al. (1987) determined the solubility of hydrogen-bearing species in aluminosilicate and albite melts at temperatures up to 1500 °C and pressures up to 25 kbar, finding dissolution of $OH^-$, $H_2O$, and $H_2$ and that $H_2$ solubility is significantly lower than that of $H_2O$ for the same melt composition (Persikov et al., 1990; Luth et al., 1987).

Hirschmann et al. (2012) found molecular $H_2$ solubility in quenched basaltic and andesitic glasses between 0.7–3 GPa to increase relative to the total amount of dissolved hydrogen (in the forms of $H_2O$, $OH^-$ and $H_2$) at lower $fO_2$ and higher total pressure. A recent study by Foustoukos (2025) used in-situ Raman spectroscopy in a hydrothermal diamond anvil cell at 0.3–1.3 GPa and 600–800 °C to determine the equilibrium distribution of molecular $H_2$ between $H_2O$-saturated sodium aluminosilicate melts and $H_2O$-bearing fluid. He identified a decrease in dissolved molecular $H_2$ with increasing temperature, suggesting that ingassing of hydrogen into a rocky planet's magma ocean may be less efficient than prior studies claimed (Foustoukos, 2025).

Molecular hydrogen is the dominant gaseous species in protoplanetary disks, for which the solar $pH_2/pH_2O$ ratio is $\sim 2 \times 10^3$ at ~1500 K. A planet forming in equilibrium with a near-infinite reservoir of nebular gas will therefore host an $H_2$-dominated atmosphere. Should the Earth, and other rocky planets, have accreted within the lifetime of the solar nebular gas (5 Myr), they would have captured quantities of $H_2$-rich atmospheres commensurate with their masses (Stökl et al., 2015). Recent works have considered whether sufficient quantities of nebular-derived $H_2$ could have dissolved into Earth's magma ocean, before undergoing reactions with molten silicates in the interior which could have constituted the source of Earth's water (e.g., Young et al., 2023). While the isotopic composition of Ne suggests it was nebular-derived (Williams and Mukhopadhyay, 2018), the isotopic ratios of hydrogen and nitrogen (D/H, $^{15}N/^{14}N$) and other noble gases (Kr, Xe) indicate instead a chondritic or cometary origin Marty (2012), Alexander et al. (2012), Mukhopadhyay and Parai (2019). In addition, the bulk concentration of H in chondrites exceeds Earth's present-day hydrogen budget (e.g., Piani et al., 2020). Therefore, while it is still unknown whether Earth ever accreted a substantial nebular atmosphere, it is likely that other sources contributed to Earth's hydrogen budget.

Olson and Sharp simulated Earth's formation using a model of magma ocean dynamics coupled to a primary atmosphere, finding that if Earth accreted to at least 30% of its final mass while the nebular gas disk was present, several ocean masses worth of H could have ingassed into Earth's interior (Olson and Sharp, 2019). Similarly, Young et al. (2023) found that nebular dissolution in progenitor planetary embryos would be consistent with the Earth's water inventory via chemical equilibrium reactions between the dissolved, nebular-derived hydrogen and the Fe–Mg–Si–O–H system (assuming a dry starting composition for the embryos) (Young et al., 2023). However, the accuracy of both models is uncertain, given the lack of constraints as to the solubility of hydrogen-bearing species in silicate liquids over the relevant pressure-temperature-composition range.

Despite existing works on H-bearing species' solubilities in silicate melts, there is still a lack of experimental data on water solubility at lower pressures (~1 bar) over range of melt compositions, $fH_2/fH_2O$ ratios and temperature. To fill this gap, here we present a series of experiments performed using a 1-bar $H_2$–$CO_2$ gas mixing furnace on a variety of silicate (anorthite-diopside-based) melts under a reducing range of $fH_2/fH_2O$ ratios (i.e., $\Delta IW$ between −4 to −1) and temperatures from 1673 to 1823 K. We discuss our results in the context of nebular ingassing as a mechanism for delivering hydrogen-bearing species to rocky planets.

## 2. Methods

### 2.1. Starting materials

The silicate melt compositions used in this study derive from O'Neill and Eggins (2002) (their Table 1) and were synthesized by mixing high-purity (> 99%) oxide and carbonate powders of MgO, $SiO_2$, $CaCO_3$, $Al_2O_3$, and $Fe_2O_3$ (Table 1). Prior to weighing, the oxide powders were fired at 1000 °C ($CaCO_3$ at 200 °C) overnight to minimize contamination from atmospheric adsorption. The powders were weighed to the desired proportions, and the resultant powder was mixed under acetone in an agate mortar for 10 min followed by a dry-down. This process was repeated three times. To decarbonate the samples, we heated them at 1000 °C in air overnight and stored the powdered mixtures in a desiccator prior to the experiment. We weighed the sample mixture before and after decarbonation. For the samples with iron, we synthesized them by adding various proportions of $Fe_2O_3$ to the anorthite-diopside eutectic composition.

### 2.2. Gas-mixing furnace experiments

To prepare a sample for an experiment, we mixed ~20 mg of sample powder with an adhesive composed of polyethylene oxide powder and distilled water into a slurry. This material was suspended onto ~1.5–3 mm diameter wire loops of 0.3 mm thickness. The Fe-free samples were suspended on Pt wires, whereas the samples with Fe were held on Re ribbon to minimize the loss of iron from the melt to the metal. Each experiment contained either two or five samples. For the experiments with two samples, the two wires were suspended on either end of a ~5–7 mm mullite bar which was attached to an alumina rod sample holder. In the case of five samples, a ring or 'chandelier' was constructed from braided platinum wires with samples suspended at evenly spaced intervals along the ring, and the ring was attached to the alumina rod sample holder (Supplementary Figure S1). In some cases, multiple samples would stick together or there was too much slurry used on a sample such that it all evaporated prior to the quench, so the total number of samples per experiment varied from one to five.

**Table 1**
Initial silicate compositions in this study prior to the gas-mixing furnace experiments, reported as weight %. Five compositions were studied in the quaternary CaO–MgO–Al$_2$O$_3$–SiO$_2$ (CMAS) system and three in the CMAS-"FeO" system. The compositions in the CMAS system are the same as those used in O'Neill and Eggins (2002) (their Table 1). The anorthite-diopside (AD) eutectic composition contains 42% anorthite, 58% diopside. The three samples in the CMAS-"FeO" system were obtained by adding 10, 20 and 30 wt% of Fe$_2$O$_3$ to the AD eutectic composition. The compositions of the quenched silicate melts after the experiments were measured by EPMA and reported in Supplementary Table S3. As discussed further in the Results section, for the Fe-bearing samples, the amount of FeO in the glass after the experiments decreased significantly from the original amount in the starting material. The final column is the optical basicity ($\Lambda$), which is a measure of the electron donor ability of the oxide ions in the silicate composition.

| Melt | CaO | MgO | Al$_2$O$_3$ | SiO$_2$ | FeO | $\Lambda$ |
| --- | --- | --- | --- | --- | --- | --- |
| AD eutectic | 24.1 | 10.6 | 15.2 | 50.1 | – | 0.62 |
| AD+Fo (15%) | 20.5 | 16.9 | 12.9 | 49.6 | – | 0.63 |
| AD+En (60%) | 15.2 | 20.6 | 9.3 | 54.8 | – | 0.61 |
| AD+Wo (140%) | 38.0 | 4.6 | 6.6 | 50.6 | – | 0.64 |
| AD+Qz (50%) | 16.2 | 6.9 | 10.0 | 66.8 | – | 0.57 |
| AD eutectic + 10 wt% FeO | 21.7 | 9.5 | 13.7 | 45.1 | 10.0 | 0.64 |
| AD eutectic + 20 wt% FeO | 19.3 | 8.5 | 12.2 | 40.1 | 20.0 | 0.66 |
| AD eutectic + 30 wt% FeO | 16.9 | 7.4 | 10.6 | 35.1 | 30.0 | 0.68 |

Experiments were performed at the Institute of Geochemistry and Petrology, ETH Zürich, in a GERO HTRV 1-bar vertical tube furnace (inner tube diameter of 40 mm, T < 1600 °C) with various inflowing gas mixtures of H$_2$ and CO$_2$, which control the oxygen fugacity and partial pressures of H$_2$, CO$_2$, CO, and H$_2$O. The purity of the H$_2$ gas is ≥99.995% ([mol/mol]) with impurities of O$_2$ ≤ 2 ppm, N$_2$ ≤ 40 ppm, and H$_2$O ≤ 5 ppm. The purity of the CO$_2$ gas is ≥99.9 vol. % with impurities of O$_2$ + N$_2$ ≤ 500 ppm, hydrocarbons (C$_n$H$_m$) ≤ 50 ppm and H$_2$O ≤ 250 ppm. For the experiments with a H$_2$/CO$_2$ ratio of 10 (5), the H$_2$ and CO$_2$ gas flow rates were 100 ml/min and 10 (20) ml/min, respectively. For the pure H$_2$ experiments, the H$_2$ flow rate was 100 ml/min while the CO$_2$ flow rate was 0 ml/min, and vice versa for the pure CO$_2$ experiments. The temperature of the furnace hotspot was calibrated by attaching a Type B thermocouple to one of the alumina rods and varying its height systematically 1 cm at a time and waiting for 5 min at each step in order for the bead to reach thermal equilibrium. This exercise was performed at three different temperatures (1200 °C, 1300 °C and 1400 °C) to assess any temperature dependence. In each case, the measured temperature was within 1–2 °C of that set. The $f$(O$_2$) during the experiment is calculated from the gas mixture which was previously verified against the iron-wüstite (IW) buffer at 1200 °C by detecting the oxidation of an Fe-wire (or the reverse reduction of FeO) through electrical resistance measurements in response to changes in the H$_2$/CO$_2$ ratio, and therefore $f$(O$_2$). The bracketed value where oxidation and reduction is observed based on the gas flow is, to within 0.07 log units, identical to the $f$(O$_2$) at IW derived by O'Neill and Powneeby (1993). For pure CO$_2$, this is likely to be better than +/-0.15 log units.

The experiments varied in duration from ~1-5 h. Samples were drop-quenched into liquid water or onto a copper plate. To confirm that neither the water from the slurry mixture for suspending the samples nor quenching into liquid water resulted in significant contaminated hydrogen in our samples, we analyzed samples that were run under pure CO$_2$ gas and quenched into water and onto a copper plate and found that they all contained negligible dissolved water (< 4 ppm H, or equivalently < 63 ppm OH). A summary of the experimental conditions and gas fugacities is given in Table 2.

### 2.3. Fourier-Transform InfraRed (FTIR) Spectroscopy

Experimental glasses were mounted in cold epoxy to minimize seepage of resin into cracks in the samples and sliced to ~0.3–1 mm thickness using a diamond wire saw in order to expose the samples on both sides of the mount. Samples were polished down to 1 μm grade on one side and then exposed on the other side (typically polished to ~55 μm grade).

The dissolved water contents in the silicate glasses were measured using a Bruker LUMOS II FTIR microscope at the Institute for Geochemistry and Petrology, ETH Zürich. The measurements were conducted in transmission mode under unpolarized light with a SiC Globar infrared source. Each measurement used a ZnSe beamsplitter and a liquid N$_2$-cooled mercury cadmium telluride (LN-MCT) detector, with a square integration area of 205 × 205 μm. 32 scans per measurement were taken with a 4 cm$^{-1}$ resolution over the spectral range from 6000 cm$^{-1}$ to 600 cm$^{-1}$. For each sample, we obtained spectra for ~10 spots spaced evenly across the sample, and the background was measured before the spots were analyzed.

We used the OH$^-$ stretching absorbance band at 3550 cm$^{-1}$ to quantify the amount of dissolved water in the glass (e.g., Stolper, 1982). Using the Bruker OPUS software to process the data, the background spectrum, which was acquired prior to measuring the spectrum of each sample, was subtracted from the sample spectrum. We then averaged the individual spectra obtained for each sample, and with this background-corrected and averaged sample spectrum, we manually applied a polynomial baseline correction with ~15-20 points. To determine the 3550 cm$^{-1}$ feature's contribution to the absorbance, we measured its peak intensity relative to the baseline within the spectral window from 3800 ± 150 cm$^{-1}$ to 2500 ± 100 cm$^{-1}$. For each sample, we applied the baseline correction and measured the peak intensity of the 3550 cm$^{-1}$ band independently two times, in order to assess the uncertainty on $I_{3550}$. Fig. 1 shows the FTIR spectra obtained for four of the anorthite-diopside eutectic samples.

The Beer–Lambert law (Eq. (1)) relates the mole fraction of OH dissolved in the glass, X(OH), to the total absorbance of the 3550 cm$^{-1}$ band produced by the OH$^-$ stretching mode, its molar mass M(OH) (kg/mol), the path length of light transmitted through the absorbing medium (i.e., the sample thickness) ($d$ in m) and the glass density ($\rho$, kg/m$^3$), by a proportionality constant which is the molar absorption cross section for that band, $\epsilon_{3550}$ (m$^2$/mol).

$$X(\mathrm{OH}) = \frac{I_{3550} M(\mathrm{OH})}{d \rho \epsilon_{3550}} \qquad (1)$$

In this equation, M(OH) is the molar mass of OH (17.008 × 10$^{-3}$ kg/mol). The glass thicknesses ($d$) varied from 2.95 × 10$^{-4}$ m to 1.13 × 10$^{-3}$ m (see Supplementary Table S2). To determine $\rho$, the glass density at the glass transition temperature, we gathered data on glass transition temperatures (T$_g$) for a variety of anhydrous silicate melt compositions (Taniguchi, 1989; Morizet et al., 2015; Richet and Toplis, 2001; Richet et al., 2000; Sossi et al., 2023; Dingwell et al., 2004; Sipp and Richet, 2002; Hansen et al., 2022; Magnien et al., 2008). We found no clear dependence of T$_g$ on the optical basicity, $\Lambda$ (Supplementary Figure S3), and therefore use the average glass transition temperature of 701 (± 53) °C for all of our samples. Using this T$_g$, we calculated the glass densities for our samples using composition data from our electron probe microanalysis (see next section), thermal expansivities and partial molar volumes from Lange and Carmichael (1987), and compressibilities from Kress and Carmichael (1991), assuming Fe$^{3+}$/∑Fe = 0 and anhydrous compositions for all samples. Given that our samples have at most ~ 300-330 ppm of dissolved OH (equivalently, ~15-20 ppm total hydrogen), our assumption of anhydrous compositions is appropriate. We also verified that varying T$_g$ by the spread observed for the anhydrous silicate melt compositions from prior studies (~150 °C) does not significantly impact the calculated glass densities.

**Table 2**

Summary of experiments performed in this study, including number of samples (N), sample compositions, temperature, duration (t), the $H_2/CO_2$ gas ratio, calculated gas fugacities (i.e., $f(O_2)$ expressed as $\Delta IW$, $f(H_2)$ and $f(H_2O)$), and the average dissolved OH concentration (ppm) in the glasses for each experiment (with the spread between samples expressed as the $1\sigma$ standard deviation shown in small font). For the experiment with one sample, the uncertainty is its internal error ($1\sigma$, as described in Section 3.2), as is given for all individual samples in Supplementary Table S2.

| Experiment | N | Melt Compositions | T (°C) | t (min.) | $H_2/CO_2$ | $\Delta IW$ | $f(H_2)$ (bar) | $f(H_2O)$ (bar) | Avg [OH] (ppm) |
|---|---|---|---|---|---|---|---|---|---|
| ADEu240723 | 2 | ADEu | 1400 | 181 | 10 | −1.86 | 0.821 | 0.088 | 208.0$_{1.1}$ |
| ADEu160823 | 2 | ADEu | 1400 | 62 | 10 | −1.86 | 0.821 | 0.088 | 208.3$_{2.4}$ |
| ADEu220823 | 2 | ADEu | 1400 | 61 | 5 | −1.19 | 0.678 | 0.156 | 256.1$_{7.1}$ |
| ADEu061123 | 2 | ADEu | 1400 | 62 | $CO_2$ | 6.65 | – | – | 35.0$_{7.2}$ |
| ADEu121223 | 2 | ADEu | 1400 | 246 | $H_2$ | −4.13 | 0.983 | 0.017 | 119.1$_{37.5}$ |
| ADEu131223 | 2 | ADEu | 1400 | 120 | $H_2$ | −4.13 | 0.983 | 0.017 | 98.8$_{1.6}$ |
| CMAS-010224-1 | 5 | AD+Fo, AD+En, AD+Wo, AD+Qz, ADEu | 1400 | 148 | 5 | −1.19 | 0.678 | 0.156 | 292.9$_{43.6}$ |
| CMAS-010224-2 | 3 | AD+En, AD+Wo, AD+Qz | 1400 | 123 | 10 | −1.86 | 0.821 | 0.088 | 221.2$_{30.2}$ |
| CMAS-020224-1 | 5 | AD+Fo, AD+En, AD+Wo, AD+Qz, ADEu | 1400 | 137 | 10 | −1.86 | 0.821 | 0.088 | 222.7$_{33.1}$ |
| CMAS-020224-2 | 1 | AD+En | 1400 | 123 | $H_2$ | −4.13 | 0.983 | 0.017 | 112.5$_{13.8}$ |
| CMAS-150224-1 | 2 | AD+Fo, AD+Qz | 1400 | 144 | $H_2$ | −4.13 | 0.983 | 0.017 | 112.9$_{1.7}$ |
| CMAS-150224-2 | 3 | AD+En, AD+Wo, AD+Qz | 1400 | 149 | $CO_2$ | 6.65 | – | – | 52.9$_{9.0}$ |
| CMAS-210224 | 3 | AD+En, AD+Wo, ADEu | 1550 | 148 | 5 | −1.24 | 0.676 | 0.157 | 274.5$_{32.3}$ |
| ADEuFeO-020524 | 2 | ADEu+20%FeO | 1400 | 120 | $H_2$ | −4.06 | 0.982 | 0.018 | 103.9$_{1.6}$ |
| ADEuFeO-220524 | 2 | ADEu+10%FeO | 1400 | 121 | $H_2$ | −4.13 | 0.983 | 0.017 | 110.0$_{13.2}$ |
| ADEuFeO-240524 | 2 | ADEu+30%FeO | 1400 | 121 | $H_2$ | −3.98 | 0.980 | 0.020 | 91.7$_{11.6}$ |
| HT1475-040724 | 2 | ADEu | 1475 | 120 | $H_2$ | −4.13 | 0.983 | 0.017 | 164.8$_{27.3}$ |
| HT1550-120724 | 2 | ADEu | 1550 | 111 | $H_2$ | −4.13 | 0.983 | 0.017 | 172.8$_{47.2}$ |

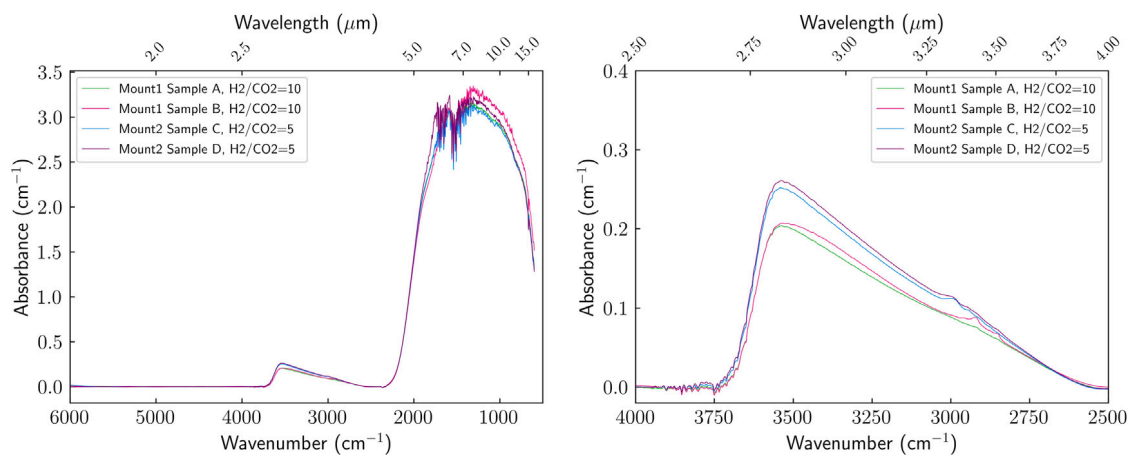

**Fig. 1.** Background-normalized and baseline-corrected FTIR spectra for four anorthite-diopside eutectic glass samples. Green and pink samples were under a $H_2/CO_2$ gas ratio of 10, while the blue and purple samples were under a $H_2/CO_2$ gas ratio of 5. (Left) Full FTIR spectra covering ~1.7-17 μm (6000–600 cm$^{-1}$) showing the OH stretching absorbance band at ~3550 cm$^{-1}$ and the silicate glass features below 2000 cm$^{-1}$. (Right) FTIR spectra of only the OH stretching absorbance band at ~3550 cm$^{-1}$, whose peak intensity ($I_{3550}$) is used to calculate the dissolved water concentration, see Eq. (1). For the FTIR spectra of all samples, see Supplementary Figure S2.

We also gathered data on the molar absorption cross sections for the 3550 cm$^{-1}$ band ($\epsilon_{3550}$) for a wide set of silicate compositions including various basalts, andesite, rhyolite, and peridotite (Dixon et al., 1995; Mandeville et al., 2002; Pandya et al., 1992; Yamashita et al., 1997; Silver and Stolper, 1989; Shishkina et al., 2014; Bondar et al., 2023; Sossi et al., 2023; Newcombe et al., 2017; Okumura et al., 2003). By comparing $\epsilon_{3550}$ to the compositions' optical basicity ($\Lambda$) (Fig. 2), we find that the molar absorptivity decreases with increasing $\Lambda$. We performed a Bayesian linear regression on the data and determined the following relationship:

$$\epsilon_{3550} = -15.84 \pm 1.13 \times \Lambda + 15.77 \pm 0.03 \quad (2)$$

Optical basicities were calculated for each sample using the composition measurements from electron-probe microanalysis (see next section) and determined $\epsilon_{3550}$ using Eq. (2). The uncertainties on all of these quantities ($d$, $\rho$, and $\epsilon_{3550}$) are propagated to those on the dissolved water content in the glass, reported as $1\sigma$. This finding suggests that for more basic, depolymerized melts (i.e., higher $\Lambda$), OH groups are less effective infrared absorbers because the local bonding environment around OH is more ionic leading to lower transition dipole moments for the O-H stretching vibration, and hence lower $\epsilon_{3550}$).

### 2.4. Glass composition measurements with electron-probe microanalysis

Compositions of the quenched glasses were determined using a JEOL JXA-8230/8350 electron probe microanalyzer (EPMA) at ETH Zürich. The EPMA is equipped with five wavelength-dispersive spectrometers. Acceleration voltage was set to 15 keV, and a 20 nA beam current, and 20 μm beam diameter. Counting time was 30 s for all elements. For Fe-bearing samples, a focused beam was used to check for possible change of composition towards the Fe metal "blobs" (see Results section for further discussion of the Fe-bearing samples). The calibration standards used were albite ($NaAlSi_3O_8$) for Si and Na, synthetic forsterite ($Mg_2SiO_4$) for Mg, anorthite ($CaAl_2Si_2O_8$) for Ca and Al, and hematite ($Fe_2O_3$) (for Fe). We obtained 10–20 analysis points per sample, distributed homogeneously across the diameter of the sample.

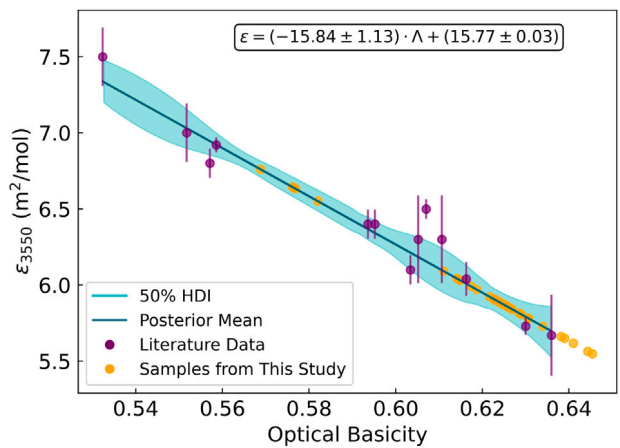

**Fig. 2.** Molar absorption cross section for the 3550 cm$^{-1}$ band ($\epsilon_{3550}$, (m$^2$/mol)) vs. optical basicity ($\Lambda$). The purple points are from the literature (i.e., Dixon et al., 1995; Mandeville et al., 2002; Pandya et al., 1992; Yamashita et al., 1997; Silver and Stolper, 1989; Shishkina et al., 2014; Bondar et al., 2023; Sossi et al., 2023; Newcombe et al., 2017; Okumura et al., 2003) and the error bars correspond to their 1$\sigma$ uncertainties. The dark teal line is the fit to the literature data determined via a Bayesian linear regression (i.e., the mean of the posterior predictive distribution, with fit values shown in the upper text box) and the light teal shaded region shows the 50% highest density interval of the posterior predictions. The orange points are the calculated $\epsilon_{3550}$ values for this study's samples using the linear fit (Eq. (2)) and the samples' optical basicities determined via their composition measurements with EPMA.

## 3. Results

### 3.1. Sample textures and compositions

All Fe-free samples quenched into homogeneous glasses with no crystal formation. For the iron-bearing samples run at $f(H_2) \geq 0.98$ bar, patches of metallic blobs were observed dispersed throughout the sample and were found to be pure Fe (Supplementary Figure S4). Consequently, the bulk composition of the glass contains significantly lower FeO contents (on average 1.13 ± 0.1 wt.%) than in the original starting mixtures (10–30 wt.%, Supplementary Table S1). Major element compositions of all glasses and their associated 1$\sigma$ uncertainties for all experiments are reported in Supplementary Table S3.

### 3.2. Gas fugacities in experiments

For experiments run in $CO_2$–$H_2$ gas mixtures, the resulting equilibrium speciation was determined by Gibbs Free Energy minimization using FactSage (Table 2) and $f(O_2)$ reported as $\log_{10}$ units relative to the iron-wüstite buffer ($\Delta IW$) from Hirschmann (2021). For the pure $H_2$ gas experiments, the $f(O_2)$ is not buffered, and depends on the local impurities in the gas. To estimate $f(O_2)$ for these runs, we used the mole fraction of FeO in the silicate and of Fe in metal in the iron-bearing samples via:

$$\Delta IW = 2 \log_{10}\left(\frac{a(FeO)}{a(Fe)}\right) \quad (3)$$

where $a(Fe)$ is 1 (pure iron metal), $a(FeO)$ is the mole fraction of FeO multiplied by its activity coefficient, as calculated using Equation 4 of Wood and Wade (2013) (cf. O'Neill and Eggins, 2002; Wood and Wade, 2013). Using this derived $f(O_2)$, we calculated the $H_2O/H_2$ fugacity ratio of the gas phase via:

$$\frac{f(H_2O)}{f(H_2)} = \frac{f(O_2)^{0.5}}{K_{eq}} \quad (4)$$

where $K_{eq}$ is the equilibrium constant for the reaction $H_2O = H_2 + 0.5\,O_2$, which is given as a function of temperature (in K) from the IVTANTHERMO database:

$$\log_{10}(K_{eq}) = \frac{-12794}{T} + 2.7768 \quad (5)$$

Therefore, in the nominally pure $H_2$ experiments and a known total pressure of 1 bar, $f(H_2)$ and $f(H_2O)$ are ~0.983 bar and ~0.017 bar, respectively. For the higher temperature experiments (1475 and 1550 °C), the speciation does not vary significantly from that at 1400 °C. Table 2 summarizes the experimental conditions, including the gas fugacities, of each set of experiments.

### 3.3. Hydrogen content and speciation in glasses

To determine the detection limit of H in our experimental glasses and to test for any significant contamination of our samples before, during and after the experiments, we performed several 'blank' experiments in 1-bar $CO_2$ gas (denoted "Pure $CO_2$" in Table 2). In these experiments, all glasses had dissolved OH contents less than 63 ppm, which is at least ~2 times lower than that measured in all of our other samples (aside from one of our samples with added iron that also had low dissolved OH content of 84 ppm). Given that the dissolved OH contents in the samples under pure $CO_2$ gas are so small compared to the measured concentrations in our other samples, the OH concentrations from Eq. (1) do not require any correction or normalization to the pure $CO_2$ experiments. We consider ~60 ppm of dissolved OH to be the detection limit of our FTIR analyses.

In all of the glasses, we detect OH$^-$, as evidenced by the prominence of the OH$^-$ stretching band at 3550 cm$^{-1}$ (2.86 μm). It is important to note that while infrared spectroscopy is sensitive to both molecular $H_2O$ and OH$^-$ absorption bands in glasses, this does not necessarily reflect the equilibrium speciation of OH$^-$ and molecular $H_2O$ in the corresponding high temperature liquids (Dingwell and Webb, 1990; Cody et al., 2020). We do not detect molecular $H_2$ (the 4130 cm$^{-1}$ band) in any of our glasses by FTIR. This observation is consistent with the results of other experiments conducted at 1 bar, including Newcombe et al. (2017) and Sossi et al. (2023), which only detect the 3550 cm$^{-1}$ OH$^-$ band. Although $H_2(g)$ does not have a permanent dipole moment (i.e., it is homonuclear), its distortion by the silicate glass structure leads to a very weak dipole that is barely detectable by infrared spectroscopy. On the other hand, $H_2$ is Raman active, specifically its stretching vibrational mode at ~4130 cm$^{-1}$, making it more sensitive to detection by Raman compared to FTIR. While we analyzed a subset of our samples using Raman spectroscopy, we nevertheless do not observe any $H_2$ dissolved in our glasses (Supplementary Figure S5).

To assess the homogeneity within a given sample, we measured the peak intensity of the 3550 cm$^{-1}$ absorption feature by FTIR for each of the ~10 individual spectra. These were averaged to determine the sample's $I_{3550}$. We assessed the individual spectra for two glass samples, F1-A-Sample3 and 121223-Sample1 which have average dissolved OH concentrations of 323 and 93 ppm, respectively. We determined that within a particular sample, variations in $I_{3550}$ correspond to standard deviations of the dissolved OH concentrations of 38 and 12 ppm, respectively, indicating that the OH is homogeneously distributed in the glasses (Supplementary Table S4).

A summary of the experimental results and the parameters used to calculate the dissolved OH content is given in Supplementary Table S2. The solubility of water (as OH$^-$) increases with $H_2O$ fugacity (Fig. 3). Using Eq. (1) and the linear fit for $\epsilon_{3550}$ (Eq. (2)), the dissolved OH contents for our silicate glasses range from 83.5 ± 12 to 334.5 ± 33 ppm.

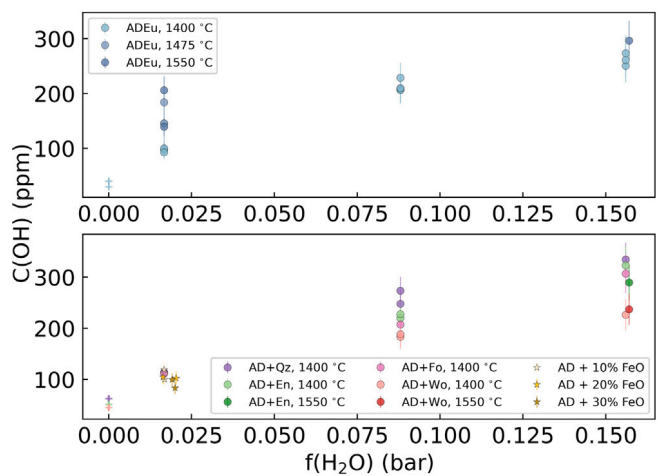

**Fig. 3.** Dissolved water content in the glasses (expressed as OH, in ppm), quantified using the height of the OH absorption stretching band at ~3550 cm$^{-1}$, as a function of the H$_2$O fugacity in the gas phase (in bar). The top figure shows all anorthite-diopside eutectic glasses, and the three shades of blue indicate the temperature of the experiments (1400, 1475, 1550 °C, with darker color indicating higher temperature). The bottom figure shows the additional samples with various CMAS starting compositions (round points) (see Table 1), excluding the AD eutectic glasses, and samples with iron (star points). The different shades of the star points correspond to the amounts of iron (added as FeO) in the starting composition (10, 20 and 30 wt%, with darker shades indicating higher amounts of added iron). In both figures, the crosses represent experiments run under pure CO$_2$.

## 4. Discussion

### 4.1. Mechanism of water dissolution

In the gas phase, the main hydrogen-bearing species, H$_2$O and H$_2$, are in equilibrium according to:

$$H_2 (g) + \tfrac{1}{2} O_2 (g) = H_2O (g) \tag{6}$$

Based on our findings that OH$^-$ is the sole detectable species in silicate melts across a range of $f(H_2)$ and $f(H_2O)$, from $f(H_2)$ of 0.68 to 0.98 bar and $f(H_2O)$ from 0.02 to 0.16 bar at 1 bar total pressure, two main sets of chemical reactions at equilibrium are required (see also Sossi et al., 2023). The first describes the equilibrium between molecular H$_2$O in the gas phase and in the melt:

$$H_2O (g) = H_2O (l) \tag{7}$$

However, given that only OH$^-$ is observed, it must then dissociate into two hydroxyl groups:

$$H_2O (l) + O^{2-} (l) = 2\,OH^- (l) \tag{8}$$

An analogous reaction can be written to describe H$_2$(g) in equilibrium with molecular H$_2$ dissolved in the melt:

$$H_2 (g) = H_2 (l) \tag{9}$$

However, as we are unable to detect H$_2$ in our experiments, we cannot place constraints on the equilibrium constant of Eq. (9). Instead, in order to estimate the solubility of molecular H$_2$ in silicate liquids at 1 bar total pressure, we make the implicit assumption that the hydrogen dissolution mechanism is H$_2$(g) = H$_2$(l) as in Reaction (9). That is, its concentration is governed by the same physical solubility mechanism as that which governs the dissolution of the noble gases (e.g., Jambon et al., 1986). Carroll and Stolper (1993) noted that the concentration of a given noble gas measured in the silicate glass decreased log-linearly with its atomic diameter, and, for a given noble gas, increased with increasing melt polymerization. Here, we use the noble gas solubility data from Carroll and Draper (1994) and three independent estimates from the hard sphere model (Chapman and Cowling, 1970), as well as the Lennard-Jones 12–6 model (Chapman and Cowling, 1970; Bird et al., 2006) to compute atomic- and molecular diameters of gas species, which are themselves derived by fitting experimental viscosity data. In each case, the log-linear trends as a function of diameter in melts equilibrated over the temperature range 1200–1400 °C at 1 bar permit the Henrian solubility constant of H$_2$ (whose diameter is slightly larger than that of Ne) to be estimated as 59(+/−5) ppm/GPa for komatiite, 177(+/−10) ppm/GPa for basalt and 537(+/−17) ppm/GPa for rhyolite. Therefore, at the highest H$_2$ fugacities of our experiments (0.983 bar), the amount of dissolved H$_2$ is below 6 × 10$^{-3}$ ppm, multiple orders of magnitude less than the amount of dissolved water (as OH$^-$).

The oxygen anions in Reaction (8) could be bridging (i.e., oxygen that links two network-forming tetrahedra cations), non-bridging (i.e., oxygen that links one tetrahedral cation to a network modifying cation) or a "free" oxygen (Newcombe et al., 2017). The hydroxide anions represent hydroxyl functional groups that are bound to ensembles of the silicate components in the melt, which we explore further in Section 4.3. Since the gas species are in equilibrium (Reaction (6)) and we detect only OH$^-$ in our glasses (Section 3.3), then, assuming this OH$^-$ represents the equilibrium, high temperature species in the melt, we need only one reaction to describe the dissolution of water vapor, which is given by combining Reactions (7) and (8):

$$H_2O (g) + O^{2-} (l) = 2\,OH^- (l) \tag{10}$$

This reaction is governed by the equilibrium constant ($K_{eq}$), given by:

$$K_{eq} = \frac{a(OH^-)^2}{f(H_2O)\,a(O^{2-})} \tag{11}$$

Eq. (11) illustrates how H$_2$O fugacity relates to its dissolution as OH$^-$ in silicate melts. Assuming an ideal solution as has been done in prior water solubility studies (e.g., Sossi et al., 2023; Newcombe et al., 2017), $a(OH^-) = X(OH^-)$. By rearranging the above expression, water solubility (in units of $X(OH^-)$) as a function of $f(H_2O)$ is given by:

$$X(OH^-) = (K_{eq}\,a(O^{2-}))^{0.5}\,f(H_2O)^{0.5} \tag{12}$$

where the equilibrium constant ($K_{eq}$) is a function of both temperature and the standard Gibbs free energy change of the reaction:

$$K_{eq} = \exp\left(\frac{-\Delta G^o}{RT}\right) \tag{13}$$

and the activity of oxygen is given by:

$$a(O^{2-}) = X(O^{2-})\,\gamma(O^{2-}) \tag{14}$$

where $\gamma$ is the activity coefficient. The amount of oxygen in the melt is related to the concentrations of the melt's various oxide components, as will be discussed further in Section 4.3. Therefore, as Eqs. (12)–(14) demonstrate, the dissolved mole fraction of the hydroxide species depends not only on the square root of the H$_2$O fugacity, but also on the temperature and melt composition. However, because silicate melts are not charged, neither hydroxide nor oxygen exist as free ions, but rather as 'functional groups' (*i.e.,* hydroxyl) bound to cations within the silicate liquid. As such, the activity coefficients of OH$^-$ and O$^{2-}$ are not readily determined in multicomponent silicate liquids, prompting a more empirical approach to be taken.

### 4.2. Deriving a water solubility law with Bayesian parameter estimation

We performed two sets of Bayesian parameter estimation in order to constrain water solubility models based on our experimental data alone and a combined dataset consisting of our data and those of Newcombe et al. (2017) and Sossi et al. (2023) (all at 1 bar; see Table 3). The models considered varied in complexity from the simplest models with no temperature or melt composition dependence, to models with a temperature-dependence only (i.e., $B(X_i)$) are treated as constants in Eq. (15)), and to the most complex models that include linear combinations of the various melt oxide parameters (Eq. (16)). We analyzed over 60 models fit to both the data presented in this study and also the combined dataset, with the number of parameters estimated ranging from one, for the simplest model, to six, for the most complex model.

Table 3
Summary of experiments performed to determine water solubility (measured as OH) in silicate melts at 1 bar total pressure and over a range of temperatures and melt compositions.

| Reference | Compositions | Temperature | $f(H_2)$ | $f(H_2O)$ | # of samples |
| --- | --- | --- | --- | --- | --- |
| This work | ADEu, CMAS, Fe+ADEu | 1673–1823 | 0.68–0.98 | 1.6E–2–0.16 | 39 |
| Sossi et al. (2023) | Peridotite | 2173 | 1.28E–5–0.064 | 7.1E–4–0.027 | 10 |
| Newcombe et al. (2017) | ADEu, Lunar Glass | 1623 | 3.24E–5–0.93 | 9.93E–3–0.32 | 25 |

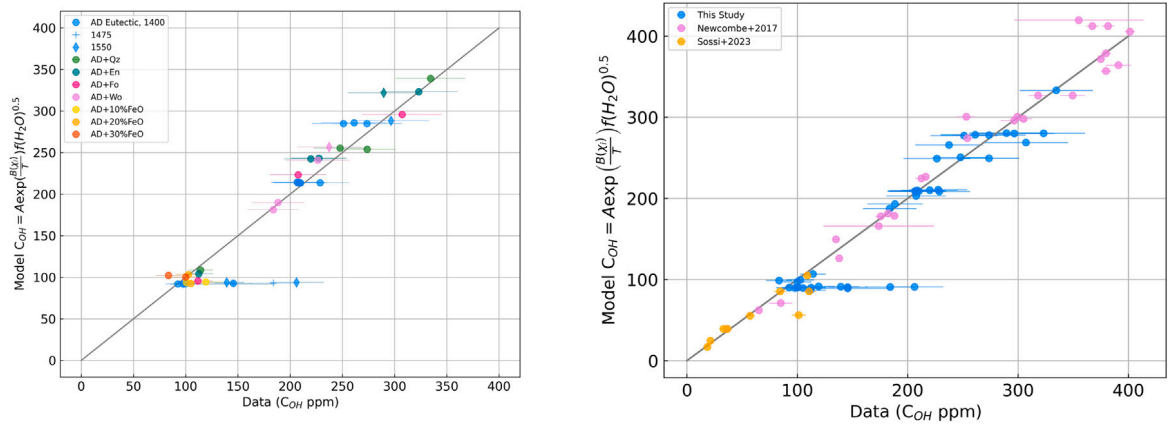

**Fig. 4.** (Left) Comparison of the dissolved OH concentrations (ppm) from our experimental data compared to the predicted concentrations according to the preferred model given by Eq. (17). (Right) Comparison of the dissolved OH concentrations (ppm) from the combined dataset including experiments from this study, Newcombe et al. (2017) and Sossi et al. (2023) compared to the predicted concentrations according to the preferred model given by Eq. (18). The horizontal errorbars in both figures show the $1\sigma$ standard deviations of the data.

$$X(OH) = A \exp\left(\frac{B(X_i)}{T}\right) f(H_2O)^{0.5} \quad (15)$$

$$B(X_i) = b_1 X_{CaO} + b_2 X_{SiO_2} + b_3 X_{MgO} + b_4 X_{Al_2O_3} + b_5 X_{FeO} \quad (16)$$

Based on Bayes' theorem, Bayesian inference involves fitting a statistical model using both available prior knowledge of the parameters for the model and then updating the parameters based on the observed data. The result from such analysis is a probability distribution for the parameters of the statistical model.

Following common practice in Bayesian parameter estimation, all predictors (i.e., $H_2O$ fugacity, temperature, and oxide mole fractions) and outcomes ($X(OH)$) were scaled by their respective means before running the regressions to ensure that all parameters are on a similar scale such that the Markov-Chain Monte Carlo (MCMC) captures the underlying data-generating distribution (McElreath, 2015). For each of our regressions, the priors are set as Gaussian (univariate normal) distributions with a mean of 0 and standard deviation of 10. The likelihood function follows the Student's t-distribution, with standard deviation given by that of our experimentally determined OH concentrations and degrees of freedom set by an exponential prior on the normality parameter. Using the Student's t-distribution for the likelihood function makes the Bayesian parameter estimation robust against outliers. A description of the quantitative comparison within each model set and how the preferred models were determined is provided in the Supplement Section 8.1. The reported parameters for the preferred models are presented in Table 4 and have physical units (e.g., bar$^{-0.5}$ and K) such that they can be used directly to calculate water solubility.

The preferred model using the experimental data from this study alone has the following functional form:

$$X(OH) = A \exp\left(\frac{b_1 X_{CaO} + b_2 X_{SiO_2}}{T}\right) f(H_2O)^{0.5} \quad (17)$$

where the fitted parameters (A, $b_1$, $b_2$) are given in Table 4. The A term has units of bar$^{-0.5}$, and $b_1$, $b_2$ have units of K. As illustrated in Fig. 4 (Left), this model sufficiently fits our experimental data with an $r^2$ value of 0.83. We note that the model does not fit the data as well for our anorthite-diopside eutectic samples under $H_2$-only gas and at various temperatures between 1400 and 1550 °C, with measured OH concentrations spanning ~90 to 200 ppm. This is likely due to the limited number of samples measured at higher temperatures and high $f(H_2)$ conditions.

Performing a similar set of Bayesian parameter estimation on the more comprehensive dataset consisting of the experiments from this study, Newcombe et al. (2017) and Sossi et al. (2023), we determined that the preferred solubility model is:

$$X(OH) = A \exp\left(\frac{b_1 X_{CaO} + b_2 X_{SiO_2} + b_3 X_{MgO} + b_4 X_{Al_2O_3} + b_5 X_{FeO}}{T}\right) f(H_2O)^{0.5} \quad (18)$$

where the means and standard deviations of the parameters' posterior distributions are given in Table 4. Of all the models considered for fitting to the combined dataset, Eq. (18) presents the preferred model, with an $r^2$ of 0.93 (Fig. 4, Right). This model includes additional compositional terms compared to the model based on our data alone due to the fact that Newcombe et al. (2017) and Sossi et al. (2023) studied melt compositions (i.e., peridotite, lunar basalt and Fe-free basalt) with more extreme (either higher or lower) MgO, Al$_2$O$_3$ and FeO contents compared to our samples so the model is better able to capture the dependence on these oxides compared to the model based on our data alone. As this model incorporates additional data from two other experimental studies (Newcombe et al., 2017; Sossi et al., 2023), we will use this water solubility law (Eq. (18), and parameter values in Table 4) for the following discussions and calculations, and we recommend that this law be used when assessing water solubilities at low-pressure conditions ($\lesssim$1 kbar) and temperatures between ~1600–2100 K. This law can be extrapolated up to ~ kbar pressures since $H_2O$ dissolution as OH has been shown to be proportional to the square-root of $H_2O$ fugacity (Hamilton et al., 1964), and our model follows this functional form. As an additional check, we compared our solubility model to the experimental findings from Shishkina et al. (2010) and find reasonable agreement (within a factor of 2) between the dissolved concentrations predicted from our model to their experimental results for a tholeiitic basaltic melt at 1250 °C at pressures up to 1 kbar.

**Table 4**
The means of the posterior distributions for the parameters in the preferred solubility models. The first row contains the parameters for the model based on the experiments in this study alone, as given by Eq. (17). The second row contains those parameters for the model based on the combined experiments from this study, Newcombe et al. (2017) and Sossi et al. (2023), as given by Eq. (18). The small values are the standard deviations of the parameters' posterior distributions. Parameter "A" is the de-scaling factor and therefore does not have an associated uncertainty.

| | A | $b_1$ (K) | $b_2$ (K) | | | |
|---|---|---|---|---|---|---|
| | 7.67E-4 | −1878.4$_{402.8}$ | 794.5$_{195.5}$ | | | |
| | A | $b_1$ (K) | $b_2$ (K) | $b_3$ (K) | $b_4$ (K) | $b_5$ (K) |
| | 7.19E-4 | −1511.1$_{513.3}$ | 886.5$_{315.9}$ | −1015.2$_{268.9}$ | 890.6$_{795.4}$ | −1755.9$_{364.5}$ |

*4.3. Effect of temperature and melt composition on water solubility*

In general, the dissolved concentration of a particular volatile species in any melt is proportional to its fugacity or its fugacity raised to some power (0.5 in the case of $H_2O$). Prior studies of volatile solubilities such as sulfur (O'Neill and Mavrogenes, 2022; Boulliung and Wood, 2022) have used this proportionality to define a capacity term that describes the temperature-, pressure- and/or melt composition-dependence on the solubility of a particular species. A capacity has the form:

$$C_Y(T, X_i, P, fO_2) = X(Y)/[f(Y)]^\varepsilon \quad (19)$$

where $C_Y$ is the capacity of species $Y$ which is directly proportional to the dissolved concentration of $Y$, $X(Y)$, and inversely proportional to its fugacity, $f(Y)$, raised to some power ($\varepsilon$). The capacity can be a function of temperature ($T$), melt composition ($X_i$), total pressure ($P$), and/or oxygen fugacity, depending on the experimental conditions varied during the study. Here, we define a "Hydrogen Capacity" term, $C_H$. Following Eq. (19), Hydrogen Capacity has the following functional form:

$$C_H(T, X_i) = A\exp\left(\frac{b_1 X_{CaO} + b_2 X_{SiO_2} b_3 X_{MgO} + b_4 X_{Al_2O_3} + b_5 X_{FeO}}{T}\right) \quad (20)$$

where $C_H$ has units of $bar^{-0.5}$ (or $ppm/bar^{0.5}$ if the dissolved concentration of OH is expressed in ppm instead of mole fraction) and is dependent on temperature ($T$), and the oxide abundances in the melt. Fig. 5 shows how hydrogen capacity varies with temperature, from 1400–2000 K, and melt composition. We find that the H capacity does not vary significantly with temperature. For a granitic melt composition, H capacity decreases slightly with temperature, but the variation is small, and the variations are even smaller for the other melt compositions. However, H capacity does vary with melt composition, increasing by nearly a factor of two from a peridotitic melt to that of granite. This variation with melt composition is primarily due to differences in the $SiO_2$ content of the melts (e.g., granite with $X(SiO_2)$ ~ 0.8 compared to the other melt compositions with ~0.4–0.5). In the context of Eq. (12), which shows that $X_{OH^-}$ is proportional to $aO^{2-}$, this dependence is unexpected, as depolymerized melts, such as those with the composition of peridotite, should have higher activities of NBOs ($O^{2-}$) than do granitic melts. Therefore, experiments over a wider range of temperatures and melt compositions are needed to confirm this behavior.

**5. Implications**

*5.1. Nebular ingassing during earth's formation*

For a growing planet embedded in an $H_2$-rich accretion disk, the partial pressure of the surrounding atmosphere increases proportionally with the mass of the planet (Perri and Cameron, 1974). To quantify this relationship, we use the dynamical modeling results from Stökl et al. (2015) who simulated the masses of nebular-accreted atmospheres on planetary bodies in 1D by solving the equations of hydrostatic equilibrium. We used their Fig. 2 to derive a relationship between the gas pressure at the surface of a molten planet and that body's mass (Supplementary Figure S6, Stökl et al., 2015). We assume that the nebular gas density is $10^{-9}$ g/cm$^3$, which has been found to be a reasonable estimate according to other studies of the midplane gas density in the solar nebula (e.g., Desch, 2007) and yields:

$$log(P_s) = 2.50(\pm 0.04)log(M_p/M_E) + 2.49(\pm 0.02) \quad (21)$$

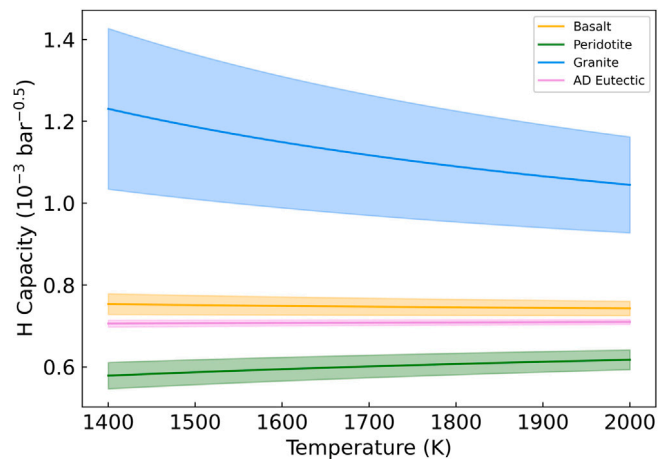

**Fig. 5.** Hydrogen capacity as a function of temperature (from 1400–2000 K) for melts of various compositions: basalt (yellow), peridotite (green), granite (blue) and anorthite-diopside eutectic (purple). The parameters for Equation (20) are reported in Table 4 and determined by the Bayesian parameter estimation discussed in Section 4.2. For each melt composition, the shaded region shows the $1\sigma$ standard deviation derived by propagating the model parameter uncertainties. In general, the H capacity does not depend strongly on temperature but does vary with melt composition.

This surface pressure-planet mass relationship predicts that a 1 $M_{Earth}$ planet would have a surface pressure of ~ 300 bars. With this relationship, we calculated the $f(H_2)$ and $f(H_2O)$ at the surface for planetary bodies with different masses, from 0.1 to 5 $M_{Earth}$, and $f(O_2)$, from $\Delta IW = -10$ to $\Delta IW = +2$ (using the IW buffer from Hirschmann, 2021), all assuming a surface temperature of 1673 K. Likely bounds on the $f(O_2)$ are $\Delta IW \sim -6.5$ set by the solar nebula at 1300 K (e.g., Grossman et al., 2008) and $\Delta IW \sim 0$, set by the present-day $Fe^{3+}/\sum Fe$ content of the Earth's mantle (Sossi et al., 2020). In these calculations, we assume that the nebular atmosphere is composed of only $H_2$ and $H_2O$, and that changes in the ratio of their fugacities does not affect the gas accretion efficiency given by Eq. (21) (though see Pahlevan et al., 2025). Using these fugacities and the $H_2O$ solubility relation derived from the experiments performed in this study along with two prior studies (Sossi et al., 2023; Newcombe et al., 2017) (Eq. (18)) and assuming a peridotite melt composition (KLB-1 from Takahashi, 1986), we estimate the amount of water (expressed as total hydrogen) that can dissolve into the magma ocean of a growing rocky planet.

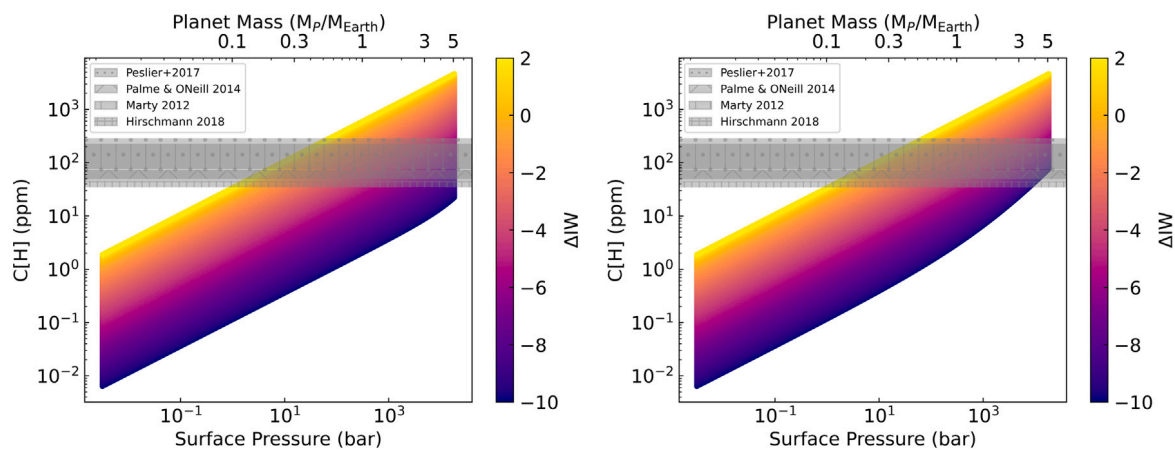

**Fig. 6.** (Left) Estimated amount of hydrogen (ppm) that would be ingassed into a planet's magma ocean using the water solubility law derived from the experiments of this study, Newcombe et al. (2017) and Sossi et al. (2023) (Eq. (18)) as a function of the planet's surface pressure (bar), mass ($M_{Earth}$), and oxygen fugacity ($\Delta$IW). (Right) Estimation of hydrogen's ingassed concentration using the sum of the water solubility law derived in this work (Eq. (18)) and the $H_2$ solubility law based on the log-linear relationship between solubility and atomic/molecular diameter observed in noble gases for komatiite melts (see text for details). In both figures, the gray horizontal bars show estimates of the total hydrogen content in the bulk silicate Earth determined by different studies (Peslier et al., 2017; Palme and O'Neill, 2014; Marty, 2012; Hirschmann, 2018).

Present-day mantle estimates of the total hydrogen content in the bulk silicate Earth range from ~40 ppm (Hirschmann, 2018; Palme and O'Neill, 2014) to ~250 ppm (Marty, 2012; Peslier et al., 2017). Comparing these estimates to the dissolved hydrogen concentrations (Fig. 6), we find that a 1 $M_{Earth}$ planet, with ~300 bar surface pressure could dissolve ~20 ppm of hydrogen at $\Delta$IW ~ −6, which is at least a factor of two less than current BSE estimates. Higher $f(O_2)$s lead to higher amounts of ingassed H due to the increasing $f(H_2O)/f(H_2)$ ratio, combined with the higher solubility of $H_2O$ with respect to $H_2$ (Fig. 6). For example, if the $f(O_2)$ of the overlying atmosphere was more oxidized at $\Delta$IW ~ −3, then ~100 ppm of hydrogen would dissolve into the molten interior, consistent with current BSE estimates.

Taking into account both the $OH^−$ (this study) and $H_2$ species in the melt (Fig. 6, Right), with $H_2$ solubility estimated via the log-linear relationship with molecular diameter as discussed in Section 4.1, the amounts of dissolved hydrogen are similar to the $OH^−$-only case for oxidized conditions above ~IW-4 and surface pressures below 1 kbar. However, at low $f(O_2)$s (<IW-4) and surface pressures above ~1 kbar, the dissolved hydrogen concentrations are greater than the $OH^−$-only case (Fig. 6, Left). This is due to the Henrian solubility relationship of $H_2$ (i.e., dissolved $H_2$ concentration being proportional to $fH_2$) enhancing hydrogen solubility at high surface pressures and under reducing conditions. For example, at ~5 kbar corresponding to a 3 $M_{Earth}$ planet, this combined solubility relation predicts about 88 ppm of dissolved hydrogen assuming $f(O_2)$ at $\Delta$IW = −6, which is 15 ppm more than estimates based on water solubility alone. This becomes more extreme at further reducing conditions, with the combined (OH and $H_2$) solubility relation predicting ~30 ppm of dissolved H at $\Delta$IW = −9, which is almost a factor of two more dissolved hydrogen than the estimates based on our water solubility law alone (~ 16 ppm). In addition, it is important to note that if the temperature of the magma ocean–atmosphere interface is higher (~3000 K), then the recent results from Foustoukos (2025) suggest that molecular $H_2$ solubility will significantly decrease by up to a factor of 20 relative to 1673 K.

If the Earth was assumed to have undergone minimal H-loss thereafter then this exercise can place constraints on the mass of the Earth during the lifetime of the solar nebula. Depending on the estimate of the H content of the BSE, Fig. 6 (right) suggests that the Earth could have been no larger than 10-50 % of its present-day mass, if the surface of the magma ocean was buffered at $\Delta$IW ~ 0 (Sossi et al., 2020). Assuming that Earth's core forms concurrently with ingassing of a nebular atmosphere and that all of the reservoirs are in equilibrium with each other (i.e., magma ocean–atmosphere, magma ocean-core), then this suggests that the core could contain significant quantities of H. Partition coefficients between Fe metal and silicate vary widely, from near unity at 5–20 GPa (Clesi et al., 2018) to 30–40 at extreme pressures and temperatures up to 60 GPa and 4600 K (Tagawa et al., 2021). For typical mean pressures deduced for terrestrial core formation, values of $D^H_{met/sil}$ around 10 seem plausible (Li et al., 2020), leading to ~1000 ppm H in the Earth's core, or 0.1 wt.%, though values up to 0.5% cannot be excluded. However, it is important to note that if the Earth contains a significant amount of nebular-derived hydrogen in the core then that would lead to a D/H fractionation which would need to be reconciled with the observed chondritic D/H at Earth's surface layers (e.g., Wu et al., 2018).

The foregoing discussion shows that the proportion of Earth's hydrogen content derived from nebular ingassing is contingent upon knowledge of its rate of growth over the nebular lifetime. Insofar as the growth rate (dm/dt) of the proto-Earth is unknown, the relevance of nebular ingassing remains equivocal. Importantly, other lines of evidence indicate that the Earth's H is unlikely to have been sourced from the solar nebula, in view of (i) the near-chondritic D/H ratios at Earth's surface (Alexander et al., 2012), (ii) the high abundance of hydrogen-bearing species in chondritic material (Wasson and Kallemeyn, 1988; Lodders and Fegley, 1998) and (iii) the single-stage Hf-W age of core formation, 34 ± 3 Myr (Kleine and Walker, 2017), which post-dates nebular dissipation. It is important to note, that these calculations do not account for any outgassing and subsequent escape of hydrogen. In order for ingassing during the magma ocean phase to be the main source of Earth's hydrogen, there would need to be minimal escape of this ingassed hydrogen and Earth's building block materials themselves could not have been enriched in hydrogen-bearing species. Treating the escape of hydrogen is beyond the scope of this study and is an avenue for future work.

### 5.2. Nebular ingassing as a source of volatiles for low-mass exoplanets

While nebular ingassing was likely not the only and may not have been the dominant source of Earth's hydrogen, it can be an important volatile delivery mechanism for low-mass exoplanets. Exoplanet population studies suggest that most rocky exoplanets captured some primary, nebular-derived atmospheres early in their histories, which may persisted for significant periods of time (e.g., Ginzburg et al., 2018). Recent works have investigated the impact of such $H_2$-rich envelopes in shaping atmospheric composition and subsequent evolution of rocky planets (Krissansen-Totton et al., 2024; Tian and Heng, 2024). The findings from our experiments combined with previous $H_2O$ (Sossi et al., 2023; Newcombe et al., 2017) and $H_2$ (Hirschmann et al., 2012) solubility studies provide ground-truth constraints on hydrogen-bearing species' dissolution into molten silicate mantles of rocky exoplanets with overlying primary atmospheres.

As Fig. 6 illustrates, oxygen fugacity at a planet's magma ocean surface strongly influences the amount of hydrogen that can dissolve into its interior. For example, by accounting for hydrogen dissolution as both OH and $H_2$ (Fig. 6 b), we find that a 1 $M_{Earth}$ planet with a magma ocean surface of $\Delta IW = 0$ can dissolve as much hydrogen as a 4 $M_{Earth}$ planet at more reducing conditions of $\Delta IW = -4$, closer to that set by a solar-type nebula. Super-Earth planets with masses up to $\sim$8 $M_{Earth}$, may harbor atmospheres with surface pressures exceeding $10^3$ bars (Fig. 6), equivalent to between $\sim$1000–6000 ppm H, or 0.9–5.4 wt % $H_2O$ equivalents. Therefore, should these planets have grown rapidly, within the lifetime of their stellar nebulae, then these planets should, at least initially, have contained substantial quantities of hydrogen (speciated as both $OH^-$ and $H_2$) in their interiors. This would have permitted storage of H over long timescales. As these planets cooled and crystallized, outgassing of some fraction of this initially dissolved H budget would have potentially led to the formation of surface oceans on these planets, given the low propensity for thermal escape due to their high escape velocities.

As observations of low-mass exoplanet atmospheres increase in the coming years, comprehensive volatile solubility laws like the one derived in this study are essential to properly interpret these observations and understand the link between their interiors and atmospheres and the importance of primary atmospheres in shaping volatile inventories.

## 6. Conclusions

Growing rocky planets in the solar nebula, should they be molten at their surfaces, must dissolve a certain quantity of hydrogen-bearing species proportional to the mass of the body. In order to constrain the quantity of H dissolved in silicate melts, we performed experiments designed to quantify water solubility in various silicate melts within the Ca–Mg–Al–Si–Fe–O system at 1 bar and temperatures from 1670–1820 K over a range of $f(H_2)/f(H_2O)$ ratios. We detect water dissolution as $OH^-$ over all studied conditions, and that solubility increases with $f(H_2O)$, with our samples reaching dissolved OH concentrations up to $\sim$335 ppm. Using Bayesian parameter estimation, a water solubility model was determined based on these experiments and those of Newcombe et al. (2017) and Sossi et al. (2023) relevant up to $\sim$100 bars and temperatures up to 2200 K. Using this solubility law, we estimate that an Earth-mass planet with surface pressure of several hundred bars due to the presence of a $H_2$-dominated primary atmosphere will dissolve $\sim$100 ppm of total hydrogen at $\Delta IW = -3$, matching current estimates for hydrogen in the bulk silicate Earth. Because $H_2O$ is more soluble than is $H_2$, similar dissolved H contents can be achieved by increasing $f(O_2)$ at lower planetary masses. While we emphasize that additional work is needed to determine how much of this dissolved hydrogen could be subsequently outgassed and escape or sequestered into the core, this result supports the idea that nebular ingassing could play an important role in sourcing hydrogen to young, rocky planets. However for Earth specifically, the chondritic D/H ratio of Earth's surface, the abundance of hydrogen-bearing species in chondritic material, and Earth's core formation age that post-dates the dissipation of the solar nebula suggest H dissolution by nebular ingassing might not have contributed significantly to Earth's present-day H budget. In addition, further experiments on hydrogen-bearing species' solubilities in the pressure range of $\sim$100-1000 bars and temperatures at and above 1600 K for primitive melt compositions are needed to comprehensively assess the dissolution of primary atmospheres into magma oceans and its importance in shaping the long-term volatile inventories of rocky worlds both within and beyond the Solar System.

## CRediT authorship contribution statement

**Maggie A. Thompson:** Writing – review & editing, Writing – original draft, Visualization, Methodology, Investigation, Formal analysis, Data curation, Conceptualization. **Paolo A. Sossi:** Writing – review & editing, Writing – original draft, Resources, Investigation, Data curation, Conceptualization. **Dan J. Bower:** Writing – review & editing, Writing – original draft, Supervision, Resources, Formal analysis. **Anat Shahar:** Writing – review & editing, Writing – original draft, Supervision, Investigation. **Christian Liebske:** Writing – review & editing, Writing – original draft, Supervision, Data curation. **Julien Allaz:** Writing – review & editing, Writing – original draft, Data curation.

## Declaration of competing interest

The authors declare that they have no known competing financial interests or personal relationships that could have appeared to influence the work reported in this paper.

## Acknowledgments

M.A.T was supported by NASA through the NASA Hubble Fellowship grant #HST-HF2-51545 awarded by the Space Telescope Science Institute, United States, which is operated by the Association of Universities for Research in Astronomy, Inc., for NASA under contract NAS5-26555. P.A.S. thanks the Swiss National Science Foundation (SNSF) via an Eccellenza Professorship (203668) and the Swiss State Secretariat for Education, Research and Innovation (SERI) under contract number MB22.00033, a SERI-funded ERC Starting Grant '2ATMO'. The authors gratefully acknowledge ScopeM for their support and assistance with the Raman spectroscopy measurements performed in this work. We greatly appreciate the constructive and insightful comments from Drs. Fabrice Gaillard, Dionysis Foustoukos, Laurent Tissandier, Sonja Aulbach and one anonymous reviewer, who all helped to improve the manuscript.

## Appendix A. Supplementary data

Supplementary material related to this article can be found online at https://doi.org/10.1016/j.chemgeo.2025.123048.

## Data availability

A link to the data and code used to generate the tables and figures in the paper are available online at the below Zenodo link:

Datafiles and Analysis Code for "Water Solubility in Silicate Melts: The Effects of Melt Composition under Reducing Conditions and Implications for Nebular Ingassing on Rocky Planets (Original data) (Zenodo)